\documentclass[journal=nalefd,manuscript=article]{achemso}

\usepackage[version=3]{mhchem} % Formula subscripts using \ce{}
\usepackage{amsmath}

\author{Shuolong Yang}
\affiliation[SLAC National Accelerator Laboratory]
{$\dag$ Stanford Institute for Materials and Energy Sciences, SLAC National Accelerator Laboratory, 2575 Sand Hill Road, Menlo Park, CA 94025, USA}
\alsoaffiliation[Stanford University]
{$\ddag$ Geballe Laboratory for Advanced Materials, Departments of Physics and Applied Physics, Stanford University, Stanford, CA 94305, USA}
\author{Jonathan A. Sobota}
\affiliation[SLAC National Accelerator Laboratory]
{$\dag$ Stanford Institute for Materials and Energy Sciences, SLAC National Accelerator Laboratory, 2575 Sand Hill Road, Menlo Park, CA 94025, USA}
\alsoaffiliation[ALS, Lawrence Berkeley National Laboratory]
{$\P$ Advanced Light Source, Lawrence Berkeley National Laboratory, Berkeley, CA 94720, USA}
\author{Dominik Leuenberger}
\affiliation[SLAC National Accelerator Laboratory]
{$\dag$ Stanford Institute for Materials and Energy Sciences, SLAC National Accelerator Laboratory, 2575 Sand Hill Road, Menlo Park, CA 94025, USA}
\alsoaffiliation[Stanford University]
{$\ddag$ Geballe Laboratory for Advanced Materials, Departments of Physics and Applied Physics, Stanford University, Stanford, CA 94305, USA}
\author{Alexander F. Kemper}
\affiliation[CCMC Group, Lawrence Berkeley National Laboratory]
{$\S$ Computational Chemistry, Materials, and Climate Group, Lawrence Berkeley National Laboratory, Berkeley, CA 94720, USA}
\author{James J. Lee}
\affiliation[SLAC National Accelerator Laboratory]
{$\dag$ Stanford Institute for Materials and Energy Sciences, SLAC National Accelerator Laboratory, 2575 Sand Hill Road, Menlo Park, CA 94025, USA}
\alsoaffiliation[Stanford University]
{$\ddag$ Geballe Laboratory for Advanced Materials, Departments of Physics and Applied Physics, Stanford University, Stanford, CA 94305, USA}
\author{Felix T. Schmitt}
\affiliation[SLAC National Accelerator Laboratory]
{$\dag$ Stanford Institute for Materials and Energy Sciences, SLAC National Accelerator Laboratory, 2575 Sand Hill Road, Menlo Park, CA 94025, USA}
\alsoaffiliation[Stanford University]
{$\ddag$ Geballe Laboratory for Advanced Materials, Departments of Physics and Applied Physics, Stanford University, Stanford, CA 94305, USA}
\author{Wei Li}
\affiliation[SLAC National Accelerator Laboratory]
{$\dag$ Stanford Institute for Materials and Energy Sciences, SLAC National Accelerator Laboratory, 2575 Sand Hill Road, Menlo Park, CA 94025, USA}
\alsoaffiliation[Stanford University]
{$\ddag$ Geballe Laboratory for Advanced Materials, Departments of Physics and Applied Physics, Stanford University, Stanford, CA 94305, USA}
\author{Rob G. Moore}
\affiliation[SLAC National Accelerator Laboratory]
{$\dag$ Stanford Institute for Materials and Energy Sciences, SLAC National Accelerator Laboratory, 2575 Sand Hill Road, Menlo Park, CA 94025, USA}
\alsoaffiliation[Stanford University]
{$\ddag$ Geballe Laboratory for Advanced Materials, Departments of Physics and Applied Physics, Stanford University, Stanford, CA 94305, USA}
\author{Patrick S. Kirchmann}
\email{kirchman@slac.stanford.edu}
\affiliation[SLAC National Accelerator Laboratory]
{$\dag$ Stanford Institute for Materials and Energy Sciences, SLAC National Accelerator Laboratory, 2575 Sand Hill Road, Menlo Park, CA 94025, USA}
\author{Zhi-Xun Shen}
\email{zxshen@stanford.edu}
\affiliation[SLAC National Accelerator Laboratory]
{$\dag$ Stanford Institute for Materials and Energy Sciences, SLAC National Accelerator Laboratory, 2575 Sand Hill Road, Menlo Park, CA 94025, USA}
\alsoaffiliation[Stanford University]
{$\ddag$ Geballe Laboratory for Advanced Materials, Departments of Physics and Applied Physics, Stanford University, Stanford, CA 94305, USA}

\title
  {Thickness-Dependent Coherent Phonon Frequency in Ultrathin FeSe/SrTiO$_{3}$ Films}

\keywords{ultrathin films, time-resolved photoemission, coherent phonons, high-temperature superconductivity\\}

\begin{document}

\begin{tocentry}

\begin{center}
\includegraphics{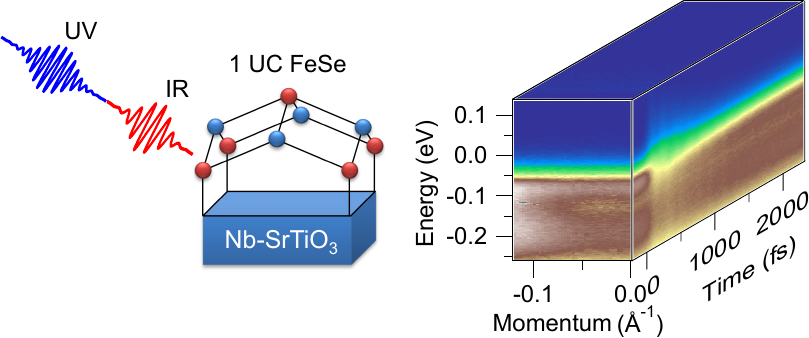}
\end{center}

\end{tocentry}

This paper is published in Nano Letters, copyright $\copyright$ American Chemical Society after peer review and technical editing by the publisher. To access the final version,

see [http://pubs.acs.org/articlesonrequest/AOR-xtHu8szuIYYXJxi6HnaD].

\begin{abstract}
Ultrathin FeSe films grown on SrTiO$_{3}$ substrates are a recent milestone in atomic material engineering due to their important role in understanding unconventional superconductivity in Fe-based materials. Using femtosecond time- and angle-resolved photoelectron spectroscopy, we study phonon frequencies in ultrathin FeSe/SrTiO$_{3}$ films grown by molecular beam epitaxy. After optical excitation, we observe periodic modulations of the photoelectron spectrum as a function of pump-probe delay for 1 unit cell, 3 unit cell, and 60 unit cell thick FeSe films. The frequencies of the coherent intensity oscillations increase from $5.00\pm0.02$ to $5.25\pm0.02$~THz with increasing film thickness. By comparing with previous works, we attribute this mode to the Se A$_\textrm{1g}$ phonon. The dominant mechanism for the phonon softening in 1 unit cell thick FeSe films is a substrate-induced lattice strain. Our results demonstrate an abrupt phonon renormalization due to a lattice mismatch between the ultrathin film and the substrate.

\end{abstract}


Ultrathin films regularly exhibit exotic electronic properties. Recent examples include the giant electron mobility in graphene~\cite{Geim2007}, and the indirect-to-direct bandgap transition in metal chalcogenides\cite{Zhang2014}. Notably, a superconducting gap opening at high temperatures was discovered in one unit cell thick FeSe films on SrTiO$_\ce{3}$ substrates (1 UC FeSe/STO)~\cite{Wang2012,Liu2012,He2013,Tan2013,Lee2014}. The gap opening temperature ($T_\textrm{c}$) of nearly $70$~K exceeds the $T_\textrm{c}$ of bulk FeSe by one order of magnitude \cite{Hsu2008}, and sets a new record for Fe-based superconductors. 1 UC FeSe represents the fundamental building block of single crystal Fe-based superconductors and can provide critical information for understanding the entire material class~\cite{Wang2012,Liu2012,He2013,Tan2013,Lee2014}.

Extensive studies on Fe-based superconductors have found that the Fe-chalcogen or Fe-pnictogen bond angle strongly impacts magnetic~\cite{Garcia-Martinez2013,Kim2012,Gerber2014,Rettig2014} and superconducting properties~\cite{Margadonna2009,Liu2011,Zhao2008,Lee2008,Kimber2009}. In FeSe/STO systems, these properties are also strongly dependent on the FeSe film thickness~\cite{Wang2012,Liu2012,He2013,Tan2013,Lee2014}. It is thus essential to study lattice properties of FeSe/STO as a function of the film thickness. 

Femtosecond time- and angle-resolved photoelectron spectroscopy (trARPES) is a powerful technique to investigate how electronic structures respond to lattice dynamics~\cite{Schmitt2008,Sobota2014,Leuenberger2013,Avigo2013,Yang2014}. With optical excitation, this technique drives coherent phonon oscillations and determines phonon frequencies with $\sim 0.01$~THz precision~\cite{Sobota2014}. In Fe-based superconductors, coherent excitation of the generic A$_\textrm{1g}$ phonon mode directly modulates the Fe-chalcogen or Fe-pnictogen bond angle~\cite{Avigo2013,Yang2014,Kim2012,Gerber2014,Rettig2014}, which can be related to transient tuning across the complex phase diagram~\cite{Kim2012}.

In this Letter, we combine a state-of-the-art trARPES setup~\cite{Sobota2012,Sobota2014} with molecular beam epitaxy (MBE)~\cite{Lee2014} to study FeSe/STO systems. We observe coherent phonon oscillations in 1 UC, 3 UC, and 60 UC FeSe films. The phonon frequency evolves from $5.00\pm0.02$~THz for the 1 UC film to $5.25\pm0.02$~THz for the 60 UC film. By comparing to a Raman scattering experiment \cite{Kumar2010}, we attribute this mode to a Se A$_\textrm{1g}$ phonon. Our phonon frequency calculation based on density functional theory (DFT) suggests that the A$_\textrm{1g}$ phonon softening in 1 UC films is predominantly due to a substrate-induced lattice strain. 

Details of the experimental setups for trARPES and MBE are discussed elsewhere \cite{Sobota2012,Lee2014}. We anneal $0.05\%$-wt Nb-doped SrTiO$_{3}$ substrates at $800^{\circ}$C for $15$~minutes. Ultrahigh purity selenium ($99.999\%$) and iron ($99.995\%$) are then evaporated to the SrTiO$_{3}$ substrates at $380^{\circ}$C, resulting in high quality thin films checked by reflection high-energy electron diffraction. The films are transferred to the trARPES system via a vacuum suitcase with a base pressure of $5\times10^{-10}$~Torr. For trARPES, we use a Coherent RegA Ti:Sapphire laser system operating at 312 kHz repetition rate. We excite the thin films using $\sim$50 fs, 1.5 eV pump pulses, and probe the ensuing dynamics by photoemitting electrons at variable time delays using $\sim$100 fs, 6 eV pulses. The energy resolution is $\sim22$~meV. For all measurements we use an incident pump fluence of $0.54$~mJ.cm$^{-2}$, and maintain an equilibrium sample temperature of $20$~K. We do not resolve dependence of the phonon frequencies on the pump fluence within the range of $0.07\sim 0.54$~$\textrm{mJ.cm}^{-2}$. DFT calculations were performed using the Quantum-Espresso package using PAW pseudopotentials~\cite{Giannozzi2009}. For comparison, calculations were done using the PBE~\cite{Perdew1996} and PBEsol~\cite{Perdew2008} functionals. We used a $11\times 11\times 7$ momentum grid; the wave function and density cutoffs were $60$~Ry and $600$~Ry, respectively.

\begin{figure*}[htp]
\begin{center}
\includegraphics[width=\columnwidth]{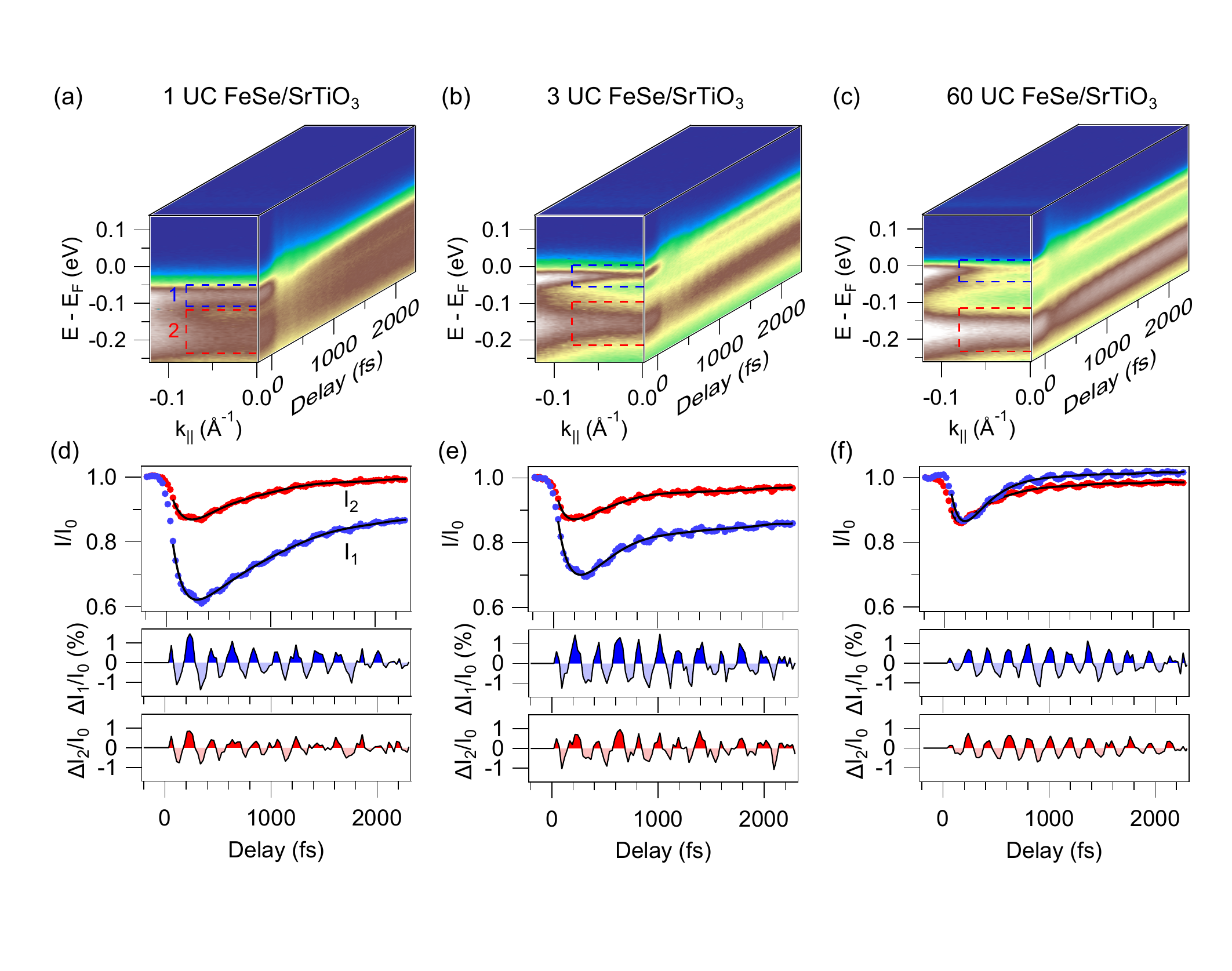}
\caption{Extraction of coherent oscillations from the trARPES data on FeSe thin films. (a)$\sim$(c) 3D visualization of the photoemission intensity as a function of energy, momentum and pump-probe delay for (a) 1 UC, (b) 3 UC, and (c) 60 UC FeSe thin films on Nb-SrTiO$_\text{3}$. We define two integration windows for Band 1 (blue) and Band 2 (red). (d)$\sim$(f) Photoemission intensity dynamics (red and blue dots) corresponding to the two integration windows as defined in (a)$\sim$(c). $I_{0}$ stands for the intensity before time zero. We use a 10th-order polynomial to account for the smooth backgrounds (solid black lines), and plot the fitting residuals which reveal oscillations with periods $\sim$200 fs.}\label{Fig1}
\end{center}
\end{figure*}

We show trARPES spectra near the Brillouin zone center for 1 UC, 3 UC, and 60 UC FeSe/STO in Fig.~\ref{Fig1}(a)$\sim$(c), where the photoemission intensity is plotted as a function of electron energy, momentum, and pump-probe delay. Before time zero, the energy-momentum cut gives an ARPES spectrum in equilibrium. For the 1 UC film, two bands are identified near $80$~meV (Band 1) and $200$~meV (Band 2) below the Fermi level ($E_\textrm{F}$). We do not observe additional features at lower binding energies as reported in the literature~\cite{Tan2013}. The observed bands are shifted up towards $E_\textrm{F}$ when increasing the film thickness from 1 UC to 3 UC, which is consistent with a previous ARPES investigation and signifies a change in the charge transfer from the substrate to the film~\cite{Lee2014}. According to the literature~\cite{Tan2013,Lee2014}, there are two hole-like bands in the spectral region of Band 1 in multi-layer FeSe films. In our data, a fine splitting near $E_\textrm{F}$ is indeed observed for 60 UC films before time zero, yet it is not observed for 3 UC films possibly due to the finite energy resolution.

At time zero, the optical excitation creates a non-equilibrium electron distribution. For all film thicknesses, the spectral peaks for Band 1 near $E_\textrm{F}$ are almost depleted near time zero, while those for Band 2 are partially depleted. During the relaxation process, periodic modulations of the spectral intensities can be observed for all three thicknesses. As shown in Fig.~\ref{Fig1}(a)$\sim$(c), we choose energy-momentum windows to be centered at these bands, and integrate the intensities within these windows. The corresponding intensity dynamics in panels (d)$\sim$(f) clearly display oscillations with periods of $\sim200$~fs (frequencies of $\sim 5$~THz). We subtract smooth backgrounds by fitting the intensity dynamics with a 10$^\textrm{th}$ order polynomial. All fitting residuals exhibit these oscillations. 

\begin{figure*}
\begin{center}
\includegraphics[width=6in]{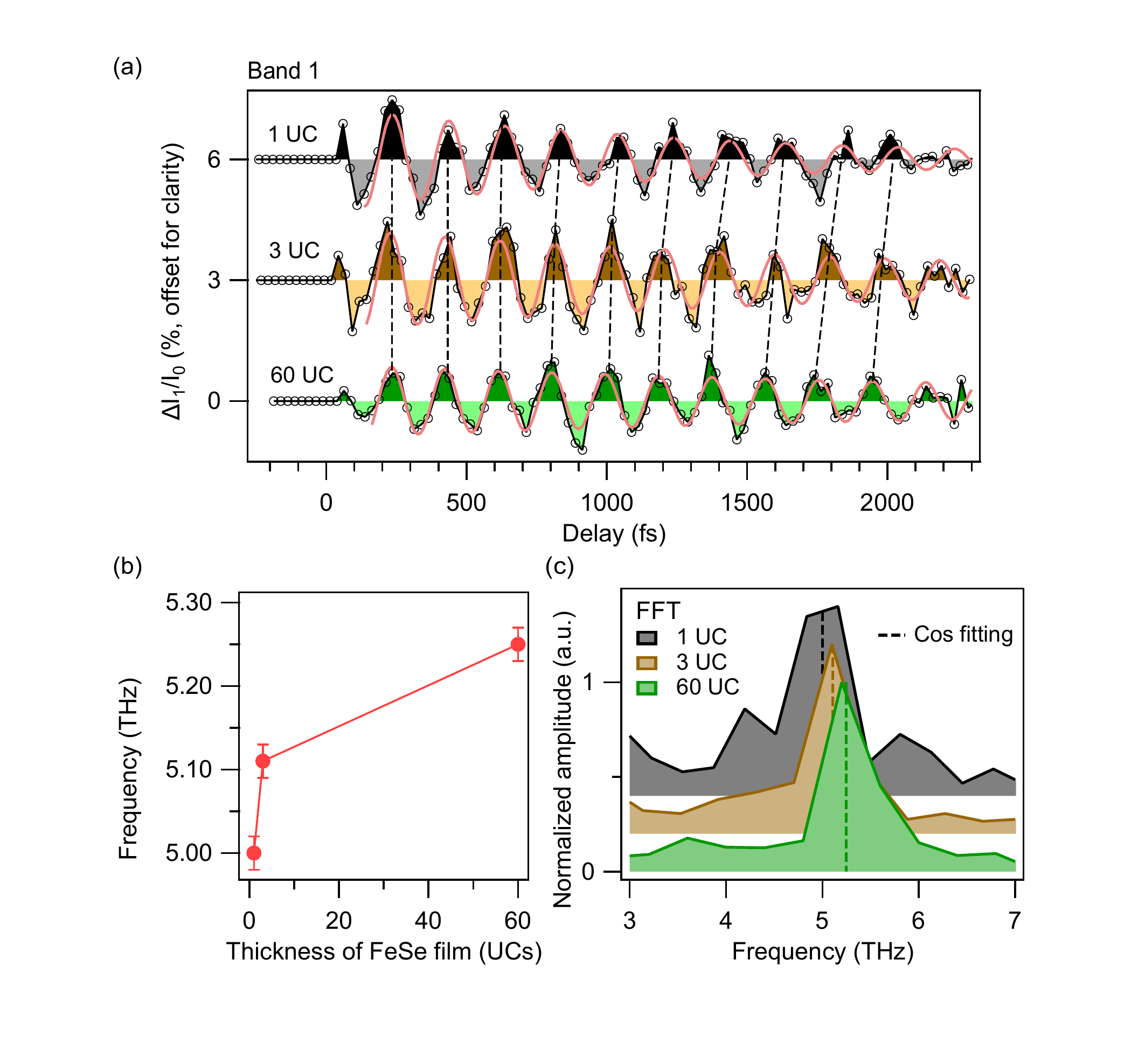}
\caption{Analysis of the coherent oscillations for Band 1. (a) Direct comparison of the fitting residuals for different film thicknesses. A frequency shift is identified by comparing the corresponding peaks of the oscillations (dashed black lines). These oscillations are fitted by a cosine function multiplied by an exponential decay envelope (solid pink lines). (b) Comparison of the frequencies extracted from cosine fitting. (c) Comparison of the Fourier transforms of the residual curves. The frequencies from cosine fitting are marked by dashed lines.}\label{Fig2}
\end{center}
\end{figure*}

We present a more detailed frequency analysis in Fig.~\ref{Fig2}. To highlight small frequency shifts, we use dashed lines to trace the oscillation peaks in Fig.~\ref{Fig2}(a). While the oscillation peaks at early delays are temporally aligned, the peaks for the 1 UC and 3 UC films at $t>1000$~fs are systematically delayed. This comparison suggests a softening of the oscillation frequency for thinner films. Quantitatively, we fit the residual traces with the functional form of $A_{0}\exp{(-t/\tau)}\cos{(2\pi ft+\phi)}$. The frequency softens from $5.25\pm0.02$~THz for the 60 UC film to $5.00\pm0.02$~THz for the 1 UC film, in agreement with the trend in the Fourier transforms of the residual traces (Fig.~\ref{Fig2}(c)). Table~\ref{Table1} further demonstrates that this softening applies to both observed bands.

\begin{table}[htp]
\caption{Comparison of phonon frequencies extracted from fitting with a modified cosine function (Fig.~\ref{Fig2}(a)) for different bands and film thicknesses}
%\centering
\begin{tabular}{c c c}
\hline\hline
Thickness (UCs) & Band 1 (THz) & Band 2 (THz) \\
\hline
1 & 5.00 $\pm$ 0.02 & 4.99 $\pm$ 0.03 \\
3 & 5.11 $\pm$ 0.02 & 5.10 $\pm$ 0.03 \\
60 & 5.25 $\pm$ 0.02 & 5.25 $\pm$ 0.02 \\
\hline
\end{tabular}
\label{Table1}
\end{table}

\begin{figure}
\begin{center}
\includegraphics[width=6in]{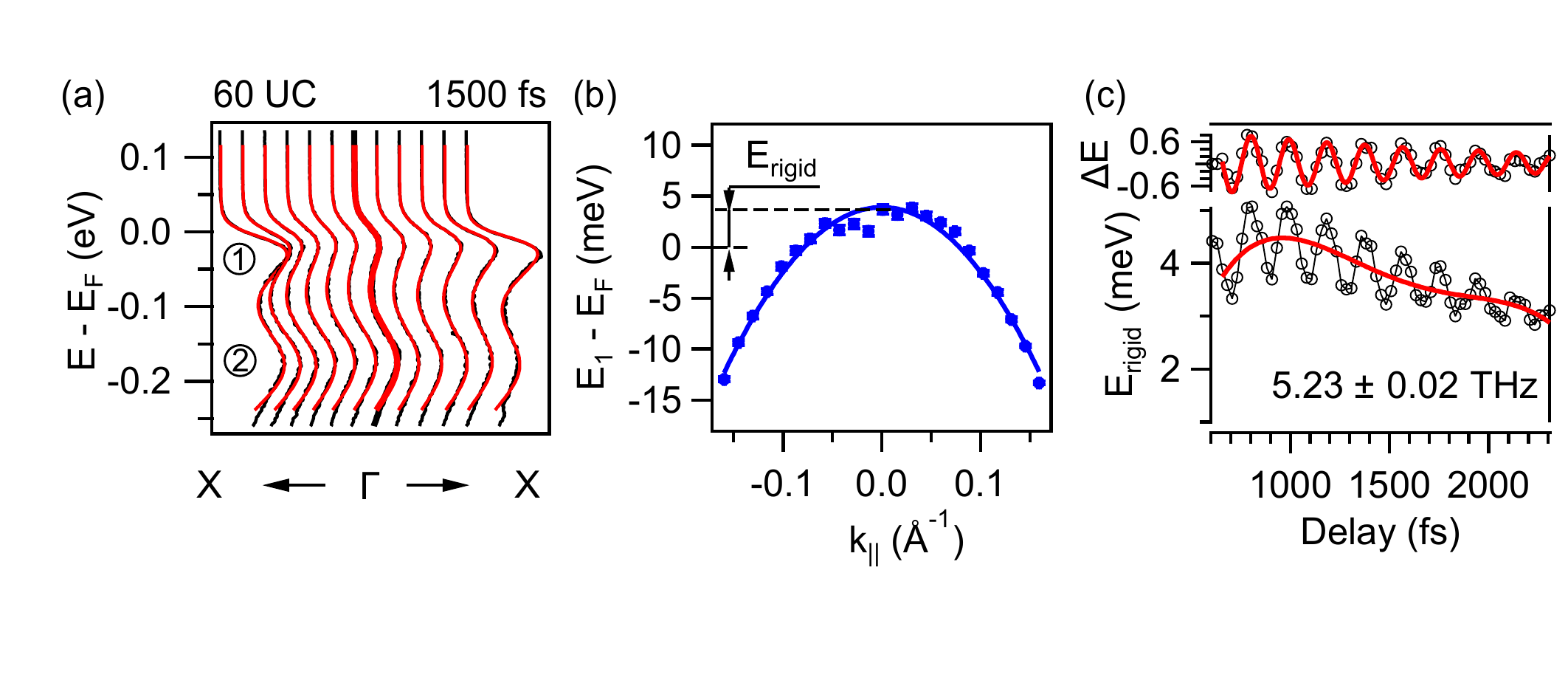}
\caption{Time-dependent dispersion analysis for 60 UC FeSe/STO. (a) Examples of energy distribution curves (black) near $\Gamma$ and the corresponding fits (red) at $1500$~fs. (b) Dispersion of Band 1 at $1500$~fs (solid circles) fitted using a parabolic function (solid curve). (c) Rigid energy shift ($E_\textrm{rigid}$, bottom hollow circles) as indicated in panel (b). The oscillatory component ($\Delta E$, top hollow circles) is extracted by removing the smooth background approximated by a 4th-order polynomial. $\Delta E$ is then fitted by a cosine function multiplied with an exponential envelope to extract the frequency of $5.23\pm0.02$~THz.}\label{Fig3}
\end{center}
\end{figure}

In addition to their intensities, the binding energies of the bands also exhibit $\sim 5$~THz oscillations. We demonstrate this using the 60 UC data, as the sharp bands observed therein permit a robust dispersion analysis. Since no band splitting is observed near $E_\textrm{F}$ after the pump excitation, we fit the energy distribution curves using two Gaussian peaks multiplied with a Fermi-Dirac distribution (Fig.~\ref{Fig3}(a)). In Fig.~\ref{Fig3}(b) we demonstrate the corresponding dispersion for Band 1 at $1500$~fs. These transient dispersions are then fitted using a parabola with a delay-dependent rigid energy shift (Fig.~\ref{Fig3}(c)). This energy shift exhibits oscillations of frequency $5.23\pm 0.02$~THz in agreement with the intensity analysis in Fig.~\ref{Fig2}.

\begin{figure}
\begin{center}
\includegraphics[width=5in]{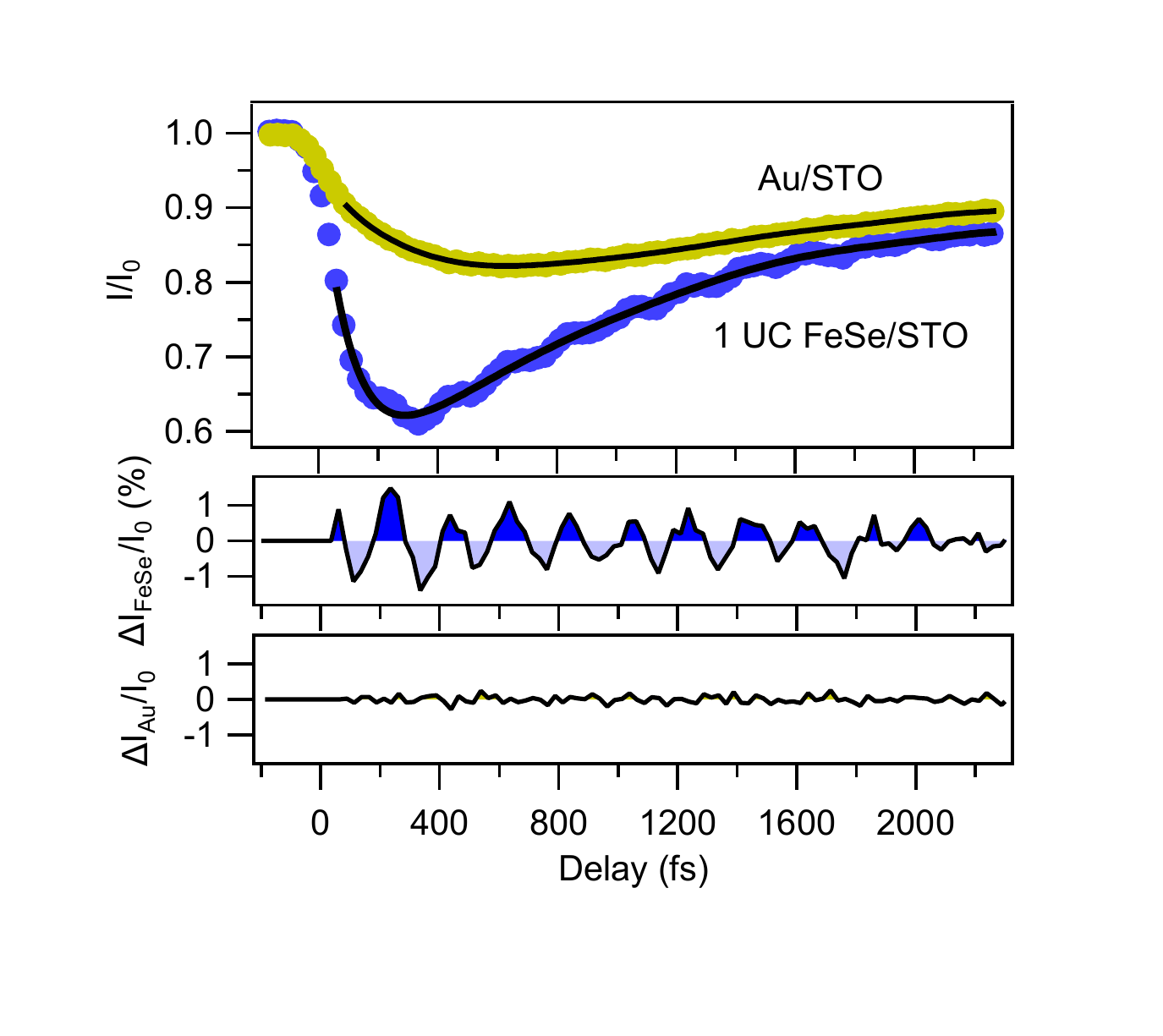}
\caption{Comparison of trARPES experiments on 1 UC FeSe/STO and Au/STO. We use a 10th-order polynomial to account for the smooth backgrounds and extract the fitting residuals (bottom plots).}\label{Fig4}
\end{center}
\end{figure}

To determine whether the $\sim 5$~THz modes originate from FeSe or STO, we perform a control experiment on a Au film of a few nm thickness on a STO substrate. As shown in Fig.~\ref{Fig4}, the data on Au/STO displays no detectable oscillations, which suggests that the $\sim 5$~THz modes on FeSe/STO are primarily from FeSe thin films. 

In accordance with a Raman scattering experiment on bulk FeSe \cite{Kumar2010}, we assign the $\sim 5$~THz modulations to the Se A$_\textrm{1g}$ mode. A$_\textrm{1g}$ modes are in general the dominant modes coherently excited by ultrafast optical excitations~\cite{Li2013}. This has been demonstrated in a variety of materials such as Bi~\cite{Leuenberger2013}, Sb~\cite{Li2013}, Bi$_{2}$Se$_{3}$~\cite{Sobota2014}, and Fe-based superconductors~\cite{Kim2012,Avigo2013,Yang2014,Gerber2014,Rettig2014}. Our observation of a dominant A$_\textrm{1g}$ mode is consistent with this general picture.

\begin{figure}
\begin{center}
\includegraphics[width=5in]{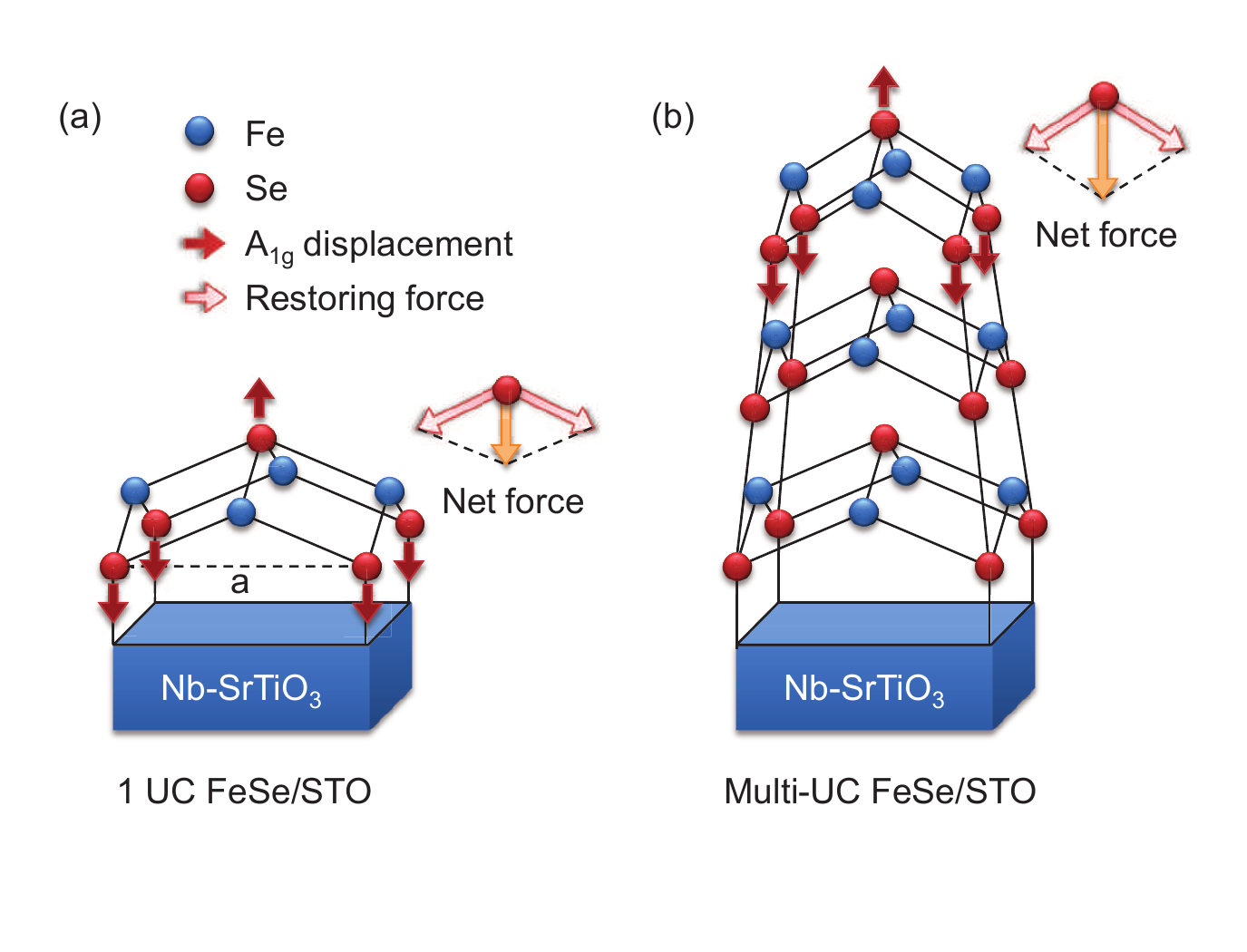}
\caption{Comparison of different film geometries impacting the A$_\textrm{1g}$ mode frequency. The Fe-Se-Fe bond angle is larger in 1 UC films, resulting in a smaller net restoring force.}\label{Fig5}
\end{center}
\end{figure}

To understand the softening of the A$_\textrm{1g}$ mode with decreasing film thickness, it is important to consider the strong coupling between the first UC of FeSe and the STO substrate~\cite{Wang2012,Liu2012,He2013,Tan2013,Lee2014}. One critical effect due to the substrate is the FeSe lattice strain~\cite{Tan2013,Li2014}. The in-plane lattice constant $a$ of FeSe increases from $3.77$~$\textrm{\AA}$ in bulk crystals~\cite{Hsu2008} to $3.9$~$\textrm{\AA}$ in 1 UC thin films matching the STO lattice constant~\cite{Tan2013}. As shown in Fig.~\ref{Fig5}, this lattice strain causes the Fe-Se-Fe bond angle to increase, which results in a lower restoring force for out-of-plane Se displacements. Here we use DFT calculations to compare the A$_\textrm{1g}$ frequencies for 1 UC FeSe with a freely relaxed lattice constant, and the one with a constrained lattice constant. We find that the lattice strain produces a $4\%$ A$_\textrm{1g}$ phonon softening, which agrees well with our experimental value of $5\%$ softening. Additional mechanisms such as interlayer interactions~\cite{Sobota2014,Zhang2011,Ishioka2015} and interface dipole field~\cite{Lee2014,Cui2015,Xiang2012} may also need to be incorporated for a complete understanding. 

Our experiment demonstrates the high potential of integrating MBE sample growth with trARPES measurements. In particular, it allows us to resolve the phonon softening of $0.25\pm 0.03$~THz ($1.0\pm 0.1$~meV) on multiple bands. This result would be difficult to obtain using traditional ARPES, and complements Raman spectroscopy by providing band sensitivity. With the capability of engineering materials on an atomic scale and studying them in non-equilibrium, this combined approach provides a platform to design and characterize electronic and lattice properties of nanoscale materials. One limitation is that the photon energy of $6$~eV does not allow us to reach the electron pocket at the Brillouin zone boundary where the superconducting gap is located~\cite{Liu2012,He2013,Tan2013,Lee2014}. This limitation can be overcome using higher energy probe photons~\cite{Rohwer2011}.

The observation of the A$_\textrm{1g}$ mode for the chalcogen or pnictogen atoms in the Fe-based superconductors has important implications \cite{Kim2012,Avigo2013,Yang2014,Gerber2014,Rettig2014}. In particular, this oscillation transiently modulates the Fe-chalcogen or Fe-pnictogen bond angle (Fig.~\ref{Fig5}), which in turn modifies the effective carrier density in the Fe $3d$ orbitals \cite{Kim2012,Yang2014}. A number of experiments have shown that the bond angle is correlated with $T_\textrm{c}$ for a variety of Fe-based superconductors~\cite{Margadonna2009,Liu2011,Zhao2008,Lee2008,Kimber2009}. It is intriguing that the observed A$_\textrm{1g}$ phonon frequency is softened in 1 UC FeSe films, which coincides with a dramatically enhanced $T_\textrm{c}$. Understanding the relationship between bond angle, phonon frequency, and electron-phonon coupling strength will be an important ingredient to identify the microscopic mechanism for high $T_\textrm{c}$ in 1 UC FeSe/STO.

\begin{acknowledgement}

The authors thank Hadas Soifer for stimulating discussions. This work was primarily supported by the U.S. Department of Energy, Office of Science, Basic Energy Sciences, Materials Sciences and Engineering Division under contract DE-AC02-76SF00515. S.-L. Y. acknowledges support by the Stanford Graduate Fellowship. J. A. S. acknowledges support from Zahid Hussain. D. L. acknowledges support from the Swiss National Science Foundation, under the Fellowship number P300P2-151328. P. S. K.'s contribution was supported in part by the National Science Foundation under Grant No. PHYS-1066293 and the hospitality of the Aspen Center for Physics.

\end{acknowledgement}

%\begin{suppinfo}
%
%This will usually read something like: ``Experimental procedures and
%characterization data for all new compounds. The class will
%automatically add a sentence pointing to the information on-line:
%
%\end{suppinfo}

\bibliography{Yang_FeSe_phonon_rev_ref}

\end{document}